\documentclass[%
 reprint,
 amsmath,amssymb,
 aps,
prb,
]{revtex4-2}

\usepackage{graphicx}
\usepackage{dcolumn}
\usepackage{bm}
\usepackage{hyperref}

\begin{document}

\preprint{APS/123-QED}

\title{Weak in the boundary: How weak SPT phases spoil anomaly matching}

\author{Daniel Sheinbaum}
 \affiliation{Division of Applied Physics, CICESE 22860, Ensenada, BC, Mexico.}
 \email{daniels@cicese.mx}
 
\author{Omar Antol\'in Camarena}
 \affiliation{Instituto de Matematicas, National Autonomous University of Mexico, 04510 Mexico City, Mexico.}
\email{omar@matem.unam.mx}

\date{\today}

\begin{abstract}
We show how weak symmetry protected topological (SPT) phases on systems with a boundary are not in 1-to-1 correspondence with weak SPT phases on fully periodic systems, breaking the standard anomaly inflow interpretation of SPT phases. We further discuss the implications for the crystalline equivalence principle (CEP).
\end{abstract}

\maketitle

\section{Introduction}\label{sec:Introduction}
The standard interpretation of symmetry protected topological (SPT) phases is that in the low energy limit we have two theories which have both an anomaly, one in the interior or bulk, and one on the edge or boundary, in such a way that the anomalies cancel out, satisfying a bulk--boundary correspondence. Weak topological phases \cite{Kane-Mele}, \cite{Fu-Kane-Weak}, \cite{Ringel-Weak} are those that arise from having discrete translation invariance instead of strong phases, which are robust to disorder. On the one hand, weak SPT phases are conventionally classified using the crystalline equivalence principle (CEP) \cite{Thorngren-Else}, which treats the discrete translation as if it were an internal symmetry. On the other hand, it is well-known that, for weak free fermion topological phases, the bulk--boundary map fails to be a 1-to-1 correspondence \cite{Thiang-Real}, \cite{Prodan-Schulz-Baldes}. We will show that the bulk--boundary correspondence must also fail to be an isomorphism for weak interacting SPT phases and propose an adapted classification scheme for weak SPT phases with a boundary. We further elucidate how there is generally a subset of weak SPT phases on systems without a boundary that do not arise when including it, breaking the standard anomaly inflow interpretation. Finally, we discuss how the existence of these phases poses problems for the interpretation of the CEP when combined with the bulk--boundary correspondence.

\section{The standard paradigm of Weak SPT phases}
Let us consider the standard picture that classifies weak SPT phases in the formalism of \cite{Thorngren-Else}, where we treat the discrete spatial translation symmetry group $\mathbb{Z}^d$ as if it consisted of internal symmetries. Then, if we also include an internal symmetry group $H$, what classifies  $d+1$-dimensional $\mathbb{Z}^d\times H$ Bosonic SPT phases i.e. weak $d+1$-dimensional $H$ Bosonic SPT phases, according to the CEP\cite{Thorngren-Else}  is the $H$-equivariant cohomology group
\begin{align}\label{eq:TE}
    H^{d+2}_{H}(T^d; \mathbb{Z}),
\end{align}
where $T^d$ is the real space unit cell $d$-dimensional torus. Note that we can obtain the strong SPT phases by restricting to the bottom-cell point of $T^d$, obtaining $H^{d+2}_{H}(pt;\mathbb{Z})$.

\section{When is the bulk--boundary map not an equivalence?}\label{sec:bbne}
There is a standard physical picture of SPT phases: they either i) have gapless edge states on their boundary as originally described by Else-Nayak \cite{Else-Nayak-SPT}, which has now been rebranded as anomaly inflow or anomaly matching \cite{Witten-Anomaly}, \cite{Freed-Hopkins-SPT}, in the original sense of 't Hooft in its modern incarnations \cite{Kapustin-Thooft}; or ii) the theory on the surface is gapped but exhibits anomalous topological order \cite{Cheng-LSM}, \cite{Wang-LSM}, \cite{Chen-LSM}. In either case the theory on the boundary of the system has an anomaly which is exactly the one required to cancel another anomaly in the other theory, higher by one dimension. The prototypical example of the former is the IQHE, where the nontrivial Hall conductance in the bulk implies a nontrivial Hall conductance at the boundary, with chiral gapless edge modes. For the latter, it can be achieved by a strong topological insulator built from a time-reversal superconductor with vortices when interactions are included \cite{Wang-LSM}. In this picture, the bulk--boundary map (that is, computing the set of SPT phases for fully periodic systems without a boundary and then mapping each phase to SPT phase of the same system with a boundary) should yield the same result, i.e. the map should be an equivalence or isomorphism. When there is indeed an equivalence this rigorously confirms the expectation that we should be able to define a phase by only looking at the bulk. However, what would happen if the symmetry defining and protecting the phase is destroyed by the boundary?
\\
Let us first consider weak SPT phases made out of free fermions since, unlike the case for interacting systems, these are classified using only basic quantum mechanical principles and do not rest upon any conjectures, making them a standard testbed for the study of topological phases. Then one can easily check that for weak free fermion SPT phases, the bulk--boundary map is not in most cases an isomorphism! There are more weak free fermion phases in fully periodic systems than in those with a boundary. We can use T-duality \cite{Thiang-Crystallographic-1} to pass from the Brilloin zone torus $\mathbb{T}^d$ to the real space torus $T^d$, so that for free fermion SPT phases on fully periodic systems are classfied by the $\mathit{K}$-theory group
\begin{align}
\mathit{KR}^{-n}(\mathbb{T}^{d}) = \mathit{KO}^{d-n}(T^{d}),
\end{align}
where $n$ is determined by the Altland-Zirnbauer tenfold way symmetry type, for example, for time reversal symmetry $\theta^2 = -I$ one has $n=4$.  On the other hand, for systems with a boundary, free fermion phases are classified by the group
\begin{align}
\mathit{KR}^{-n+1}(\mathbb{T}^{d-1}) = \mathit{KO}^{(d-1)-n+1}(T^{d-1}),
\end{align}
where $T^{d-1}$ is real space unit cell torus on the boundary. The reason for reformulating well-known results using T duality and the real torus is two-fold: first, it highlights the difference between bulk and boundary, since in both cases we are computing the same kind of $\mathit{K}$-groups $\mathit{KO}^{d-n}$ and the only difference is the unit cell ($T^d$ in the bulk case and $T^{d-1}$ in the boundary case). Second, since the interacting case is formulated in terms of the real space torus, this reformulation will guide us as to how to correctly modify the classification for weak SPT phases on systems with boundary. An explicit example that violates the standard anomaly matching picture arises in the experimentally relevant example of topological insulators in class $AII$ i.e. $n = 4,\, d = 3$ (see Theorem 3.8 in \cite{Thiang-Real}). There is an extra $\mathbb{Z}_{2}$ phase which becomes trivial by the boundary breaking the discrete-translation symmetry in said direction. Also note that when we restrict to the bottom-cell point in both cases we obtain exactly the same $\mathit{K}$-groups, namely $\mathit{KO}^{d-n}(pt)$, meaning that strong phases are the same independently whether the system has a boundary or not, as expected.\\

Here we point out that this already breaks the standard anomaly matching paradigm of SPT phases! It does so by having a bulk theory that is not reflected on the boundary. On one side this seems almost physically trivial as the boundary breaks the discrete translation symmetry in one direction so the phase is no longer protected by the former symmetry. On the other side, this reflects that boundaries can destroy/trivialize SPT phases, contrary to the standard lore \cite{Mcgreevy-Review}.

\section{The return of the bulk--boundary map for interacting weak SPT phases}
Now, how do we modify the existing classifications for weak interacting SPT phases? Let's assume Kitaev's conjecture \cite{Kitaev-SPT}, which tells us that some (as of yet unknown) cohomology theory $D_{H}$ classifies $d$-dimensional SPT phases. There are several proposals for the choice of $D_{H}$ such as topological gauge theory \cite{Wen-SPT},\cite{Kitaev-SPT}, invertible topological field theories \cite{Freed-Hopkins-SPT}, and more \cite{Gaiotto-SPT}, \cite{Shiozaki-SPT}. Our argument applies to any of these and any other possible choice of cohomology theory, we thus keep the abstract notation. One can show that for systems without a boundary, weak SPT phases are classified by the group $D^{d}_{H}(T^d)$ \cite{ADKPSS-FTI}. For systems with a boundary, we can extend Kitaev's conjecture and postulate that there is some other cohomology theory $\bar{D}_{H}$ which classifies SPT phases. Then, via exactly the same reasoning as for systems without a boundary, weak SPT phases are classified by the group $\bar{D}^{d}_{H}(T^{d-1})$. We note that the $d$ in the degree of the group is as expected since those are $d$-dimensional systems. For strong phases these groups should agree, meaning that we have the equality
\begin{align}
 D_{H}^d(pt)= \bar{D}^{d}_{H}(pt),
\end{align}
Furthermore, using that weak phases are built from lower dimensional strong phases and $D^{d}_{H}$ and $\bar{D}^{d}_{H}$ must agree for any $d$,  we see that these cohomology theories must be the same on any torus, i.e. for systems without a boundary weak SPT phases are classified by $D_H^{d}(T^d)$ where as for systems with boundary they are classified by $D_{H}^{d}(T^{d-1})$.

For the real space unit-cell torus, the bulk--boundary map is simply obtained from an embedding of $T^{d-1}$ into $T^d$ that is consistent with the choice of boundary,
\begin{equation}
 i: T^{d-1} \rightarrow T^d,
\end{equation}
which produces the desired map in cohomology
\begin{equation}\label{coh-bb-map}
    i^{*}: D_{H}^{d}(T^d)\rightarrow D_{H}^d(T^{d-1}).
\end{equation}

Using a trick well known to topologists it is easy to see which phases are destroyed/trivialized by the presence of a boundary. No cohomology theory $D_{H}$ can distinguish between a torus $T^d$ and its James splitting \cite{Hatcher-Alg-Top},\cite{ADKPSS-FTI} which, simply put, implies the following equality:
\begin{equation}
    D^{d}_{H}(T^{d-1} \times T^1) = D^d_{H}(T^{d-1}) \oplus D^{d-1}_{H}(T^{d-1}).
\end{equation}
 Thus, the kernel of the bulk--boundary map $i^*$, i.e. the set of weak phases which become trivialized when we add a boundary, is given by:
\begin{equation}
    ker\,\, i^{*} =  D^{d-1}_{H}(T^{d-1}).
\end{equation}

Notice that $i^{*}$ is onto, so all boundary phases are in the image of $i^{*}$ and should be detectable via their boundary edge modes in the spirit of Else-Nayak \cite{Else-Nayak-SPT}, Freed-Hopkins invertible TQFTs \cite{Freed-Hopkins-SPT} or via surface topological order \cite{Wang-LSM}, \cite{Chen-LSM}. Finally we point out the example of the $d=3$ free fermion weak topological insulator bulk phase in symmetry class $AII$, which was destroyed by the boundary as discussed in section \ref{sec:bbne}. This phase was shown to be stable under interactions in the invertible TQFT classification of Freed-Hopkins \cite{ADKPSS-FTI}, yet it will also become trivial in the presence of a boundary. One can see this in the computation of $D_{H}^{3-1}(\mathbb{T}^{3-1}) = \mho_{Pin^{\tilde{c}+}}^{3-1+2}(\mathbb{T}^{2}) \cong \mathbb{Z} \oplus \mathbb{Z}_2$ (where the $\mathbb{Z}$ summand corresponds simply to the analogue of the number of bands) in corollary 3.4 of \cite{ADKPSS-FTI}).

\section{The bulk--boundary map \& the CEP}

Our derivation in the previous section does not rely on the CEP
and thus applies in situations where the CEP does not hold (such
as the case of free fermions ---see \cite{SA-CEP}). When the CEP
does hold, how does it interact with the bulk--boundary map described
above? Anomaly matching for internal symmetries is equivalent to the
bulk--boundary map being an isomorphism. If the CEP holds, then
anomaly matching should also be an isomorphism for crystalline
symmetries, however the bulk--boundary map is not in general an
equivalence for weak SPT phases (and we computed the kernel of the map
to be weak SPT phases in one dimension lower).

Note that the bulk--boundary map (\ref{coh-bb-map}) explicitly breaks
the translation symmetry in one direction (which is encoded
topologically in the inclusion of tori, \(T^{d-1} \hookrightarrow T^{d}\),
used to construct the map). Thus, on the other side of the CEP, we
must either accept there is a ``special'' kind of internal symmetry which
is broken by the bulk--boundary map, or we must accept that the CEP is
not an equivalence carrying one bulk--boundary map to the other ---in
the language of category theory, one would say the CEP is not a
natural transformation. The first alternative would signal a
difference between internal symmetries arising from spatial symmetries
and ``true'' internal symmetries, suggesting the word ``equivalence'' in
the term CEP is overblown. The latter alternative merely limits the range of applicability of the CEP as an equivalence.\\
For theories which do satisfy a CEP (such as equation (\ref{eq:TE})) the bulk--boundary map must be forcibly reinterpreted as a map that \textit{explicitly} breaks the internal symmetries, $\mathbb{Z}^d\rightsquigarrow \mathbb{Z}^{d-1}$, in the same manner as has been studied for other symmetry breaking transformations \cite{Wang-Explicit}, \cite{Prakash-Explicit}, but at the cost of its unnaturality as mentioned above. However, in situations where the CEP does not hold one can in principle still construct cohomology theories for interacting SPT phases (see section V of \cite{SA-CEP}) which fit the known classification for internal symmetries, and our result applies to these as well since we only used the rigorous results of \cite{ADKPSS-FTI} in combination with the bulk--boundary map and a slight extension of Kitaev's conjecture. Thus, our results transcend the CEP.

\section{Discussion}
As we have shown above, weak SPT phases do not in general follow the physical picture of anomaly matching between the bulk and boundary theories. There are nontrivial weak SPT phases in the bulk (given by the kernel of the bulk--boundary map) that are trivialized in the presence of a boundary and break the standard anomaly inflow matching interpretation \cite{Witten-Anomaly}, \cite{Mcgreevy-Review}, \cite{Chen-LSM}, \cite{Wang-LSM}. This stems from the simple fact that the boundary breaks the discrete translation symmetry that protected them; however we have further quantified how many phases the boundary trivializes via the kernel of the bulk--boundary map, in a uniform manner regardless of the choice of cohomology theory classifying weak SPT phases, as weak phases in one dimension lower. As for the CEP, our results transcend it and do not require its use. Furthermore, we have showed that the following two statements are incompatible: i) The bulk--boundary map is an isomorphism for internal symmetries and ii) Spatial $\mathbb{Z}^d$-SPT phases are equivalent to SPT phases for an internal symmetry group. In this sense, internal and spatial SPTs are inequivalent in the presence of a boundary.

\acknowledgements{We thank G.C. Thiang for pointing us to references and examples.}


\providecommand{\noopsort}[1]{}\providecommand{\singleletter}[1]{#1}%

\end{document}